\documentclass[%
 reprint,
 amsmath,amssymb,
 aps,
prd,
nofootinbib
]{revtex4-1}

\newcommand\ee{\end{equation}}
\newcommand\be{\begin{equation}}
\newcommand\eea{\end{eqnarray}}
\newcommand\bea{\begin{eqnarray}}
\newcommand{\bV}{\mathbf{V}}

\newcommand{\HH}{\mathcal{H}}

\newcommand{\mnras}{MNRAS}

\newcommand{\jcap}{JCAP}

\usepackage[normalem]{ulem} 
\usepackage{subfigure} 
\usepackage{graphicx}
\usepackage{dcolumn}
\usepackage{bm}
\usepackage{amsmath,amsfonts,amssymb,slashed,upgreek,color}
\usepackage{multirow}
\usepackage{cancel}
\usepackage{ulem}
\usepackage[citecolor=blue]{hyperref}
\usepackage{comment}
\usepackage{multirow}
\usepackage{xcolor}

\newcolumntype{C}[1]{>{\centering\arraybackslash}p{#1}}

\begin{document}

\title{Disentangling modified gravity from a dark force with gravitational redshift}

\author{Sveva Castello$^{1}$}
\email[]{sveva.castello@unige.ch}

\author{Zhuangfei Wang$^{2}$}
\email[]{zwa168@sfu.ca}

\author{Lawrence Dam$^{1}$}
\email[]{lawrence.dam@unige.ch}

\author{Camille Bonvin$^{1}$}
\email[]{camille.bonvin@unige.ch}

\author{Levon Pogosian$^{2}$}
\email[]{levon@sfu.ca}

\affiliation{$^{1}$D\'epartement de Physique Th\'eorique and Center for Astroparticle Physics,
Universit\'e de Gen\`eve, Quai Ernest-Ansermet 24, CH-1211 Gen\`eve 4, Switzerland}
\affiliation{$^{2}$Department of Physics, Simon Fraser University, Burnaby, British Columbia, V5A 1S6, Canada}

\date{\today}

\begin{abstract} 
The standard approach to test for deviations from general relativity on cosmological scales is to combine measurements of the growth rate of structure with gravitational lensing. In this study, we show that this method suffers from an important limitation with regard to these two probes: models of dark matter with additional interactions can lead to the very same observational signatures found in modified gravity and vice versa. Using synthetic data of redshift-space distortions, weak lensing, and cosmic microwave background, we demonstrate that this degeneracy is inevitable between modifications of gravity and a dark fifth force. We then show that the coming generation of surveys, in particular the Square Kilometre Array, will allow us to break the degeneracy between such models through measurements of gravitational redshift. Performing a Markov chain Monte Carlo analysis of the synthetic dataset, we quantify the extent to which gravitational redshift can distinguish between two representative classes of models, generalized Brans-Dicke (modified gravity) and coupled quintessence (fifth force).
\end{abstract}

\maketitle

\section{Introduction} \label{sec:introduction}
The standard $\Lambda$ Cold Dark Matter ($\Lambda$CDM) cosmological model consists of two elusive components, dark energy and dark matter, whose properties remain unknown. Dark energy, which is postulated to drive the observed accelerated expansion of the Universe \cite{SupernovaSearchTeam:1998fmf, SupernovaCosmologyProject:1998vns}, takes the form of a cosmological constant $\Lambda$ within the standard model and accounts for approximately 70\% of the total cosmic density \cite{Planck:2018vyg}. The lack of understanding of its properties has motivated the search for alternative explanations of the accelerated expansion, most notably the possibility that there are modifications to general relativity (GR) on scales larger than individual galaxies. In many popular modified gravity scenarios, this is attributed to the presence of a new dynamical degree of freedom mediating the gravitational interaction, e.g.~a scalar field in the so-called scalar-tensor theories \cite{Horndeski:1974wa, Deffayet:2011gz}, or a vector field in Einstein-Aether theories~\cite{Jacobson:2000xp,Carroll:2004ai,Jacobson:2007veq}. 

In addition to dark energy, a range of observations have strongly suggested the presence of another obscure component, dark matter, based on its gravitational influence on visible matter and light. In the standard cosmological model, dark matter takes the form of a slow-moving (``cold''), noninteracting, nonbaryonic species. Although CDM has proved remarkably successful in explaining structure formation on large scales, its observational status on small scales is less clear~\cite{Bullock:2017xww,Salucci:2018hqu}. These issues, together with the fact that dark matter has yet to be directly detected, have motivated a plethora of alternatives and extensions to the simple CDM paradigm. For example, dark matter has been proposed to interact weakly with particles of the Standard Model~\cite{Barkana:2018lgd,Schewtschenko:2015rno}, with the dark sector (e.g.~dark radiation)~\cite{Diamanti:2012tg,Buen-Abad:2015ova}, or through self-interactions via a fifth (nongravitational) force~\cite{Spergel:1999mh,Pettorino_2012,Costa:2014pba,Archidiacono:2022iuu}. 

To test for gravity modifications and dark matter interactions, the standard approach is to employ measurements of the growth rate of cosmic structure obtained through redshift-space distortions (RSD) (see e.g.~Refs.~\cite{Blake:2011rj,Howlett:2014opa,BOSS:2016wmc,eBOSS:2020yzd, Nguyen:2023fip}). Since gravity governs the way cosmic structure evolves with time, deviations from GR generically modify the growth rate. Similarly, the growth rate would also be affected by any new force acting on the dark matter particles, as this would have an impact on structure formation. 

A modification of gravity would involve a deviation in the Poisson equation affecting the evolution of all cosmic components, whereas a fifth force would only modify the Euler equation for dark matter. Current analyses are testing these two scenarios separately. The community working on gravity modifications usually assumes that dark matter is a cold noninteracting particle and uses the growth rate of structure to test theories beyond GR, see e.g.~Ref.~\cite{DES:2022ccp}. By contrast, the community working on nonstandard dark matter models typically assumes the validity of GR and uses the growth rate to constrain the strength of dark matter interactions~\cite{Pourtsidou:2013nha,Buen-Abad:2017gxg,Bottaro:2023wkd}. As we do not know which scenario (if any) is correct, it is important to ask whether it is possible to test for both kinds of modifications at the same time and disentangle them from one another. 

In Ref.~\cite{Bonvin:2022tii}, it was shown that this cannot be achieved with current observables. The growth rate of structure is affected by the modifications in both scenarios, potentially in the very same way. On the other hand, weak lensing, which is sensitive to the sum of the two gravitational potentials describing the geometry of the Universe, is also generically unable to discriminate between the two cases. While it is possible to look for observational signatures of specific models (see e.g.~Refs.~\cite{Joyce:2014kja,Burrage:2018zuj, Kading:2023hdb}), these two scenarios cannot be disentangled without knowing \textit{a priori} which kind of modifications (if any) are present in the data. 

Reference~\cite{Bonvin:2022tii}, however, pointed out that the degeneracy can be broken by considering measurements of an observable accessible by the coming generation of galaxy surveys: gravitational redshift~\cite{Sobral-Blanco:2021cks, Sobral-Blanco:2022oel}. This effect, originally predicted by Einstein~\cite{Einstein:1916vd}, directly probes the gravitational potential encoding the distortion of time and entering the Euler equation. This yields an immediate test of modified gravity and provides a way to distinguish it from the presence of a fifth force acting on dark matter. 

In this work, we quantify the capability of gravitational redshift to break the degeneracy between gravity modifications and a dark fifth force, performing a Markov chain Monte Carlo (MCMC) analysis on synthetic data, based on the survey specifications of the Square Kilometre Array Phase 2 (SKA2)~\cite{Bull:2015lja}. We consider two specific models: the symmetron modified gravity model~\cite{Hinterbichler:2010es}, which is a scalar-tensor theory of generalized Brans-Dicke (GBD) type, and a coupled quintessence (CQ) model~\cite{Amendola:2016saw, Barros:2018efl} with exactly the same form of the coupling and the potential as the symmetron, but with only dark matter coupled to the scalar field. We show that the parameters of the two models are fully degenerate in an analysis including RSD, weak lensing, and cosmic microwave background (CMB) data. We then demonstrate that including gravitational redshift into the analysis breaks the degeneracy, and we determine how large the deviations from $\Lambda$CDM need to be for this observable to be effective. 

The rest of this paper is organized as follows. In Sec.~\ref{sec:two-scenarios}, we present the GBD and CQ models considered in our analysis, highlighting their degeneracies in RSD and weak lensing data. We then describe the galaxy clustering, weak lensing and CMB observables included in the analysis in Sec.~\ref{sec:observables} and discuss the details of the numerical investigation in Sec.~\ref{sec:methods}. Finally, we present the results in Sec.~\ref{sec:results} and conclude in Sec.~\ref{sec:conclusions}. We include some additional details on the GBD and CQ models in Appendix \ref{app:GBD_and_CQ} and on the specifications for the galaxy clustering observables in Appendix \ref{app:specifications}.

\section{Two degenerate scenarios}\label{sec:two-scenarios}

\subsection{GBD and CQ}
In order to illustrate the disentangling power of gravitational redshift, we compare a modified gravity model with a scenario involving a dark fifth force acting on dark matter. The argument that follows is fully general and applicable to any model belonging to these categories, but for the sake of concreteness, we focus on a GBD scalar-tensor theory and a CQ model, following Ref.~\cite{Bonvin:2022tii}. Both scenarios involve an additional scalar field, according to the actions given in Appendix \ref{app:GBD_and_CQ}. In the GBD case, the scalar field has the same conformal coupling to all cosmic components, whereas in CQ the coupling only involves dark matter (in the form of CDM) and propagates a dark force of nongravitational origin. We denote the coupling strengths with $\beta_{1}$ and $\beta_{2}$ and the scalar field masses with $m_{1}$ and $m_{2}$ in the GBD and CQ models, respectively. 

We work within linear perturbation theory and assume that the Universe is described by a perturbed flat Friedmann-Lema\^itre-Robertson-Walker (FLRW) metric in the conformal Newtonian gauge, with line element
\begin{equation}\label{eq:perturbed_FLRW}
    \mathrm ds^2 = a^2[-(1+2 \Psi)\mathrm d \tau^2 + (1- 2 \Phi)\mathrm d \mathbf{x}^2]\, ,
\end{equation}
where $\tau$ denotes conformal time and $a$ the scale factor. The two metric potentials $\Psi$ and $\Phi$ encode perturbations in the geometry of the Universe. The matter content can be described by two fields: the density contrast $\delta = \delta \rho / \rho$ and the velocity divergence $\theta$. These contain contributions from both CDM ($c$) and baryons ($b$), $\rho \delta  = \rho_b \delta_b + \rho_c \delta_c$ and $\rho \theta = \rho_b \theta_b + \rho_c \theta_c$. The relations among these fields, listed in Appendix \ref{app:GBD_and_CQ}, are provided by the chosen theory of gravity and the energy-momentum conservation. In particular, we notice that the GBD scenario involves a nonzero anisotropic stress and modifications in the Poisson equation, whereas the CQ case only includes modifications in the Euler equation for CDM. 

\subsection{Impact on the growth of cosmic structure}\label{sec:impact_structure_growth}
Let us analyze the implications of the GBD and CQ scenarios on the growth of cosmic structure. In either scenario, combining the Einstein and conservation equations yields an equation for the growth of density fluctuations in the linear regime of the form
\begin{equation}\label{eq:evol_eq_Geff}
    \ddot{\delta}+ {\cal H} \dot{\delta} = 4 \pi G_{\rm eff} a^2 \rho\delta\, .
\end{equation}
Here, overdots indicate derivatives with respect to conformal time, $\cal H$ is the conformal Hubble parameter and $G_{\rm eff}$ is the effective Newton constant that encodes the effective gravitational coupling. The latter takes the following form in the two models:
\begin{align}
 G_{\rm eff}^{\rm GBD} &= G \left[1 + {2 \tilde{\beta}_1^2k^2\over a^2m_1^2 + k^2} \right]\, , \label{eq:Geff_GBD}\\
  G_{\rm eff}^{\rm CQ} &=  G \left[1 + {2 \tilde{\beta}_2^2k^2\over a^2m_2^2 + k^2} \left( \rho_c \over \rho \right)^2\left( \delta_c \over \delta \right) \right]\, ,\label{eq:Geff_CQ}
\end{align}
where $G$ is the Newton constant and we have defined $\tilde{\beta}^2_i = \beta^2_i / (8 \pi G)$, for $i = 1,2$. We remark that the only difference between the two expressions is the term $\left( \rho_c / \rho \right)^2 \left( \delta_c / \delta \right)$ suppressing the value of $G_{\rm eff}$ in the CQ case, due to the fact that the coupling only affects CDM in this scenario. However, such a difference can be absorbed into the unknown value of the coupling $\beta_{2}$. Therefore, measurements of the growth of cosmic structure that only constrain $G_{\rm eff}$ cannot distinguish between a scenario where $\beta_1=0$ (GR is valid) and $\beta_2\neq 0$ (CDM experiences a fifth force), and a scenario where $\beta_1\neq 0$ (gravity is modified) and $\beta_2=0$ (there is no dark fifth force). In general, if we analyze the data allowing for both $\beta_{1}$ and $\beta_{2}$ to vary, we are sensitive to $G_{\rm eff}=G_{\rm eff}^{\rm GBD}+G_{\rm eff}^{\rm CQ}$. Assuming $m_1 = m_2 \equiv m$, this yields
\begin{equation}
    G_{\rm eff} = G \left[1 + {2 k^2\over a^2m^2 + k^2} \left(\tilde{\beta}_1^2 +  \left( \rho_c \over \rho \right)^2\left( \delta_c \over \delta \right) \tilde{\beta}_2^2 \right) \right],\label{eq:Geff_tot}
\end{equation}
indicating that measurements of the growth of structure can only constrain the combination $\tilde{\beta}_1^2 + \left( \rho_c / \rho \right)^2 \left( \delta_c / \delta \right) \tilde{\beta_2}^2$. While the shape of the degeneracy depends on the specific models considered, deviations in the Poisson and the Euler equations are generically indistinguishable through measurements of the growth of structure \cite{Castello:2022uuu}.

\subsection{Impact on weak lensing}\label{sec:oservables_WL}

The other key large-scale structure observable is weak lensing, measured in cosmic shear, magnification of high-redshift galaxies, and the lensing of the CMB. Weak lensing is directly sensitive to the Weyl potential, i.e.~the sum of the two gravitational potentials $\Phi$ and $\Psi$ \cite{Dodelson:2003ft}. 
However, as shown in Ref.~\cite{Bonvin:2022tii}, measurements of the Weyl potential cannot disentangle GBD from CQ, as in both scenarios we have the following Poisson-like relation,\footnote{This assumes that $A^2 \approx 1$ in the GBD action in Eq.~\eqref{eq:GBD_action}, which is required for the screening mechanism to be effective~\cite{Joyce:2014kja}. As discussed in Appendix \ref{app:GBD_and_CQ}, this assumption has no impact on the arguments presented.}
\begin{align}
\label{eq:Weyl}
k^2(\Phi+\Psi)=-8\pi G \rho\delta\, .   
\end{align}
This shows that the Weyl potential is directly related to the evolution of density fluctuations and is consequently subject to the same degeneracy found with measurements of the growth of structure. Note that modified gravity models other than GBD can lead to modifications in Eq.~\eqref{eq:Weyl} and consequently can be distinguished from a dark fifth force through measurements of weak lensing. However, the opposite statement is not true: models with a dark fifth force always leave Eq.~\eqref{eq:Weyl} unchanged and thus cannot be distinguished from a GBD modification of gravity using weak lensing. 

\subsection{Setup for the analysis}
In the following, we will show that the two types of modifications can be disentangled using gravitational redshift, which provides a direct measurement of the potential $\Psi$ appearing in the Euler equation. For the purpose of this demonstration, we adopt the symmetron model~\cite{Hinterbichler:2010es, Brax:2012gr,Hojjati:2015ojt,Wang:2023tjj}, in which the modifications become important only at late cosmological times, above a given value of the scale factor $a_{\star}$. The time evolution of $\tilde{\beta}_{1,2}$ and $m_{1,2}$ in this model is given by
\begin{align}
\tilde{\beta}(a) &= \beta_{\star} \sqrt{1-\left(\frac{a_{\star}}{a}\right)^{3}}\, , \label{eq:beta_time_evol} \\
m(a) &= {m_{\star}} \sqrt{1-\left(\frac{a_{\star}}{a}\right)^{3}} \, ,
\end{align}
for $a > a_{\star}$. We fix $a_{\star} = 0.5$ for both scenarios and study the constraints on the parameters $\beta_{*1}$ and $\beta_{*2}$ for different values of the masses that set the Compton wavelength of the scalar field. For the purpose of our demonstration, we choose a scenario where the two kinds of modifications have exactly the same time evolution, thus allowing us to treat both on an equal footing.  

\section{Observables}\label{sec:observables}

We consider three observables to constrain the GBD and CQ models: galaxy clustering, weak lensing, and the CMB.

\subsection{Galaxy clustering}

Spectroscopic galaxy surveys provide a measurement of the galaxy number counts fluctuations, 
\be
\Delta(\mathbf{\hat{n}}, z)\equiv\frac{N(\mathbf{\hat{n}},z)-\bar N(z)}{\bar N(z)}\, ,
\ee
where $N$ is the number of galaxies in a pixel centered in direction $\mathbf{\hat{n}}$ and at redshift $z$ and $\bar N$ denotes the average number of galaxies inside the pixel. The observable $\Delta$ can be expressed within linear perturbation theory as \cite{Bonvin:2011bg, Challinor:2011bk, Yoo:2009au}
\begin{equation} 
\begin{split}
\Delta(\mathbf{\hat{n}}, z)&=b\!\:\delta-\frac{1}{\HH}\partial_r(\bV\cdot\mathbf{\hat{n}})+\frac{1}{\mathcal H}\partial_r\Psi+\frac{1}{\mathcal H}{\dot \bV}\cdot \mathbf{\hat{n}} \\
&+\left(1-5s+\frac{5s-2}{\mathcal H r}-\frac{{\dot{\HH}}}{\mathcal H^2}+f^{\rm evol}\right)\mathbf V\cdot \mathbf{\hat{n}} \, , \label{eq:Delta_rel}
\end{split}
\end{equation}
where $r$ denotes the comoving distance to the galaxies, $b$ is the galaxy bias, $s$ is the magnification bias and $f^{\rm evol}$ is the evolution bias. The dominant contribution to $\Delta$ arises from the first two terms, which respectively encode the effect of matter density perturbations and RSD \cite{Kaiser:1987qv}. These are the only two contributions that are measurable with current data. The other terms are relativistic corrections suppressed on subhorizon scales by a factor of ${\HH}/{k}$, including the Doppler terms and the gravitational redshift effect given by the radial derivative of $\Psi$.
\footnote{Note that $\Delta$ is also affected by other relativistic effects suppressed by $(\HH/k)^2$ and by gravitational lensing, whose impact is negligible in the redshift range considered in this work \cite{Jelic-Cizmek:2020pkh, Euclid:2021rez}.}

We can extract information from $\Delta(\mathbf{\hat{n}}, z)$ by measuring its two-point correlation function $\xi \equiv \langle \Delta(\mathbf{\hat{n}}, z) \Delta(\mathbf{\hat{n}}', z') \rangle$. The density and RSD terms in Eq.~\eqref{eq:Delta_rel} generate three even multipoles in the correlation function: a monopole, a quadrupole and a hexadecapole, which probe the growth of structure and are the key quantities measured in a standard RSD analysis. Relativistic corrections to these three multipoles have been shown to be negligible, being suppressed by $(\HH/k)^2$ relative to the density and RSD contributions~\cite{Jelic-Cizmek:2020pkh}. One can however exploit the antisymmetry generated by relativistic effects. This is manifested in the presence of odd multipoles, most notably a dipole, when cross-correlating two different populations of galaxies~\cite{Bonvin:2013ogt,Bonvin:2014owa, Gaztanaga:2015jrs,McDonald:2009ud, Yoo:2012se,Croft:2013taa}. Since the dipole is sensitive to the effect of gravitational redshift, which is suppressed by only a single power of $\HH/k$, measuring it can help to disentangle modified gravity from a dark fifth force, as we show below.

The cross-correlation between two populations of galaxies with different luminosities---a bright and faint sample labeled B and F, respectively---generates the following even multipoles in the flat-sky approximation: 
\begin{align}
\label{eq:monopole}
        \xi_0^{\mathrm{BF}}(z,d) &= \frac{1}{2 \pi^2} \int \mathrm{d} k \, k^2 \bigg[b_{\mathrm{B}} b_\mathrm{F} P_{\delta \delta} + \frac{1}{5} \left(\frac{1}{\mathcal{H}}\right)^2 P_{\theta \theta}  \nonumber\\
        &\quad 
        - \frac{1}{3}  (b_\mathrm{B}+b_\mathrm{F}) \left(\frac{1}{\mathcal{H}}\right) P_{\delta \theta}  \bigg] j_0(kd) \,,  \\[8pt]
\label{eq:quadrupole}
\xi_2^{\mathrm{BF}}(z,d) &= -\frac{1}{2 \pi^2} \int \mathrm{d} k \, k^2 \bigg[- \frac{2}{3} (b_\mathrm{B} + b_\mathrm{F}) \left(\frac{1}{\mathcal{H}}\right) P_{\delta \theta} \nonumber\\
        &\quad  + \frac{4}{7} \left(\frac{1}{\mathcal{H}}\right)^2 P_{\theta \theta} \bigg] j_2(kd)\,, \\[8pt]
\label{eq:hexadecapole}
        \xi_4^{\mathrm{BF}}(z,d)&= \frac{1}{2 \pi^2} \int \mathrm{d} k \, k^2 \left[\frac{8}{35} \left(\frac{1}{\mathcal{H}}\right)^2 P_{\theta \theta}\right] j_4(kd)\,.
\end{align}
Here, $j_\ell$ denotes the spherical Bessel function of order $\ell$, and $P_{\delta \delta}, P_{\theta\theta}$, and $P_{\delta\theta}$ are the auto- and cross-power spectra, which depend on $k$ and $z$. The velocity divergence $\theta$ is defined in Fourier space by $\mathbf{V}=i(\mathbf{k}/k^2) \, \theta$. The dipole is given by
\begin{align} \label{eq:dipole}
& \xi_1^{\mathrm{BF}}(z,d) = -\frac{1}{2 \pi^2} \int \mathrm{d} k \, k^2 \left[\frac{(b_\mathrm{B} - b_\mathrm{F})}{\mathcal{H}}  \!\! \left(k P_{\delta \Psi}  -  \frac{P_{\delta \dot{\theta}}}{k} \right)\right. \nonumber\\
& \left. -  (b_\mathrm{B} \alpha_\mathrm{F} - b_\mathrm{F} \alpha_\mathrm{B}) \frac{P_{\delta \theta}}{k}  + \frac{3}{5} (\alpha_\mathrm{B} - \alpha_\mathrm{F}) \frac{P_{\theta \theta}}{\mathcal{H} k}  \right] j_1 (kd) \nonumber \\
& + \frac{1}{5 \pi^2} \int \mathrm{d} k \, k^2  \frac dr \left(\frac{1}{\mathcal{H}}\right) (b_\mathrm{B} - b_\mathrm{F}) P_{\delta \theta} (k, z) j_2 (kd)\,, 
\end{align}
where as a shorthand we have defined
\begin{equation}\label{eq:alpha}
    \alpha_{\mathrm{B}, \mathrm{F}} \equiv 1 - 5 s_{\mathrm{B}, \mathrm{F}} + \frac{5 s_{\mathrm{B}, \mathrm{F}} -2}{r \mathcal{H}} - \frac{\dot{\mathcal{H}}}{\mathcal{H}^2} + f_{\mathrm{B}, \mathrm{F}}^\mathrm{evol}\,. 
\end{equation}
The first term in Eq.~\eqref{eq:dipole}, proportional to $P_{\delta \Psi}$, is the contribution from gravitational redshift. This key term is not present in the even multipoles and provides the information needed to disentangle the GBD and CQ scenarios. The last term in Eq.~\eqref{eq:dipole} is a wide-angle correction, which is needed for a consistent treatment of the dipole \cite{Bonvin:2015kuc}.\footnote{Note that wide-angle corrections also enter the even multipoles but can be neglected since they are more suppressed compared to those in the dipole~\cite{Reimberg:2015jma,Bonvin:2013ogt}.} Note that antisymmetry can only be probed by cross-correlating distinct populations, meaning that the dipole signal is nonzero only for $\mathrm{B}\neq\mathrm{F}$.

\subsection{Weak lensing}

Weak lensing can be measured through cosmic shear. Here, we consider both shear-shear correlations and shear-clustering correlations (also called galaxy-galaxy lensing), both of which can be written in terms of the density power spectrum $P_{\delta\delta}$ by employing Eq.~\eqref{eq:Weyl}. We can safely neglect the covariance between the lensing observables and the spectroscopic galaxy clustering sample, since the lensing correlations are largely insensitive to the small radial modes from which the growth of structure and gravitational redshift are measured \cite{Taylor:2022rgy}.

\subsection{CMB}

We consider the CMB temperature and polarization angular power spectra (TT, TE and EE). We are interested in scenarios in which the deviations from $\Lambda$CDM appear at late cosmological times, leaving the physics at the time of last scattering unchanged. Therefore, the role of the CMB observables in this analysis is primarily to constrain the standard cosmological parameters. 

In addition to this, there are also the secondary CMB anisotropies. In particular, we note that late-time modifications to the growth of cosmic structure can impact the CMB on large scales through the integrated Sachs-Wolfe (ISW) effect. Moreover, the temperature and polarization spectra we employ in this analysis are affected by gravitational lensing. Since both ISW and gravitational lensing depend on the Weyl potential, we can use Eq.~\eqref{eq:Weyl} to relate these terms to the density power spectrum $P_{\delta\delta}$, thus providing constraints on $\beta_{*1}$ and $\beta_{*2}$. However, this does not give any contribution in discriminating between the two types of modifications, and, for this reason, we do not include separate measurements of the CMB lensing potential.

\section{Numerical analysis}\label{sec:methods}

\subsection{The general approach}
In order to quantify the degeneracy between the GBD and CQ scenarios and assess the ability of gravitational redshift to break it, we use {\sc mgcamb}~\cite{Zhao:2008bn,Hojjati:2011ix,Zucca:2019xhg,Wang:2023tjj}, a modified gravity patch of the Boltzmann code {\sc camb}~\cite{Lewis:1999bs,Howlett:2012mh}. We employ {\sc Cobaya}~\cite{2019ascl.soft10019T,Torrado:2020dgo} to carry out an MCMC analysis of the mock dataset consisting of the observables described in Sec.~\ref{sec:observables}.

We generate mock data for a given fiducial model where only one kind of modification is present, for example CQ with $\beta_{*2}=1$ and a given mass $m_{*2}$. However, when we perform a fit of the data, we treat both $\beta_{*1}$ and $\beta_{*2}$ as free parameters, fixing $m_{*1}=m_{*2}$. This does not mean that both modifications are likely to occur at the same time in our Universe, but it is the correct procedure to properly quantify the degeneracy between the two models and determine whether the input fiducial model can be recovered. The resulting constraints will either indicate the presence of a degeneracy or the lack thereof: if the two configurations $\{\beta_{*1} =0, \beta_{*2}\neq 0\}$ and $\{\beta_{*1}\neq 0, \beta_{*2}=0\}$ provide an equally good fit, we can conclude that the data cannot distinguish between the two scenarios. On the other hand, if the case $\{\beta_{*1}\neq 0, \beta_{*2}=0\}$ is excluded, the two kinds of modifications can be disentangled from one another, breaking the degeneracy.

For each mock data vector built according to the specifications in Sec.~\ref{sec:data_vector}, we perform two types of analysis to assess the degeneracy between the two models: one where we do not include the dipole in the clustering data in Eq.~\eqref{eq:data_clustering}, i.e.~there are no constraints from gravitational redshift, and another one where the dipole is included. We build a multivariate Gaussian likelihood for the combination of the three observables and adopt wide uniform priors on the free parameters. 

In each analysis, we have 11 free parameters: 5 cosmological parameters, with fiducial values $\omega_b=0.02242$, $\omega_c=0.12$, $h=0.677$, 
$A_s=2.105\times10^{-9}$, and $n_s=0.9665$; 4 bias parameters related to the galaxy populations (defined in Appendix~\ref{app:specifications}); and 2 parameters associated with GBD and CQ, $\beta_{*1}$ and $\beta_{*2}$, see Eq.~\eqref{eq:beta_time_evol}. Since the mass of the scalar field is strongly degenerate with the coupling and thus cannot be separately constrained~\cite{Wang:2023tjj}, we fix it to the same value in both scenarios, $m_{*1}=m_{*2}$, and run our analysis for different choices of the Compton wavelength $\lambda_* \equiv 1/m_*$. 

\subsection{Data vector specifications} \label{sec:data_vector}

\subsubsection{Galaxy clustering}
The galaxy clustering data vector consists of the following eight multipoles:
\begin{equation}
\mathbf{D}
=\big(\xi_0^\mathrm{BB}, \xi_0^\mathrm{BF},
    \xi_0^\mathrm{FF}, \xi_1^\mathrm{BF},
    \xi_2^\mathrm{BB}, \xi_2^\mathrm{BF},
    \xi_2^\mathrm{FF}, \xi_4^\mathrm{TT} \big)\, . \label{eq:data_clustering}
\end{equation}
This includes the usual even multipoles of RSD and importantly, also the dipole $\xi^\mathrm{BF}_1$.
Note that we only consider the hexadecapole in the total population of galaxies (labeled T), as Eq.~\eqref{eq:hexadecapole} does not contain a dependence on population-specific biases and thus considering different galaxy samples does not provide additional information. We compute the multipoles in Fourier space from power spectra generated using {\sc mgcamb} and perform the transform to configuration space using the FFTLog method (Fast Fourier Transform with Logarithmically spaced periodic sequence)~\cite{Hamilton:1999uv}. 

Since the dipole is too small to be measured from current data~\cite{Gaztanaga:2015jrs}, we perform the forecasts assuming a SKA2-like survey, for which the dipole is expected to reach a signal-to-noise ratio of 80~\cite{Castello:2023zjr}. We consider the specifications of Ref.~\cite{Bull:2015lja} for the number density, sky coverage, and galaxy bias. The separations considered in this work range from $d=20\,h^{-1}$Mpc to $d=160\,h^{-1}$Mpc, an interval where nonlinear corrections were found to be negligible for the dipole~\cite{Bonvin:2023jjq}. Furthermore, we assume a bias difference of 1 between the bright and faint galaxy populations, in agreement with that measured in Ref.~\cite{Gaztanaga:2015jrs}. More details on the modelling of the galaxy bias and the magnification bias can be found in Appendix~\ref{app:specifications}. 

We use the data vector covariance calculated in Appendix C of Ref.~\cite{Bonvin:2018ckp}, which includes both shot noise and cosmic variance. We account for the covariance between different separations and different multipoles, but neglect the one between different redshift bins, since the bins are quite wide and do not overlap thanks to the spectroscopic precision of the redshift measurements.

\subsubsection{Gravitational lensing}
For gravitational lensing, we consider the specifications of the Dark Energy Survey (DES) Year 1~\cite{DES:2017myr, DES:2017tss} and compute the shear-shear correlations and the galaxy-galaxy lensing correlations using {\sc mgcamb}. Since {\sc mgcamb} is based on linear perturbation theory, we impose the ``aggressive'' scale cut implemented in Ref.~\cite{Zucca:2019xhg} to effectively remove the data in the nonlinear regime. This is more restrictive than the cut used in the DES analysis for the $\Lambda$CDM model~\cite{DES:2017myr}, since in our case it would be incorrect to follow the same approach and assume the validity of GR to model the observables on mildly nonlinear scales. All nuisance parameters, including intrinsic alignment, lens photo-z shift, source photo-z shift, and shear calibration are fixed according to the DES Year 1 standard values~\cite{DES:2017myr, DES:2017tss}.

Note that by the time SKA2 data will be available, more precise measurements of weak lensing will have been performed by Euclid\footnote{\url{https://www.esa.int/Science_Exploration/Space_Science/Euclid}} and LSST\footnote{\url{https://www.lsst.org/}}. However, we do not include these in our forecasts, since weak lensing cannot break the degeneracy between GBD and CQ, as discussed in Sec.~\ref{sec:oservables_WL}. 

\subsubsection{CMB}
For the CMB power spectra, we consider the Planck specifications introduced in Ref.~\cite{Rocha:2003gc}, using the 143 GHz channel parameter values to compute the noise of the measurement~\cite{Zhao:2008bn}. The cosmic variance is computed according to Refs.~\cite{White:1993jr, Ng:1995nh}. 

\section{Results}\label{sec:results}

\subsection{Breaking the degeneracy}
\begin{figure}[t]
    \centering
    \includegraphics[width=0.5\textwidth]{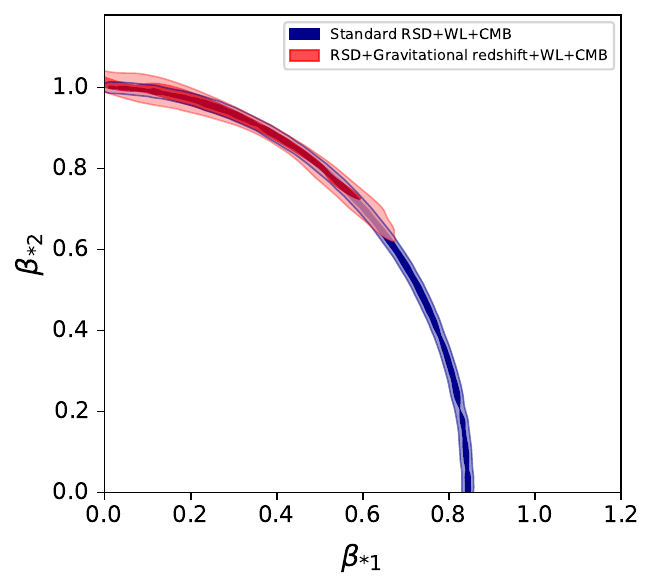}
    \caption{Marginalized $1\sigma$ and $2\sigma$ contours on $\beta_{*1}$ and $\beta_{*2}$, with and without gravitational redshift. Here, the fiducial model is CQ and is specified by $\beta_{*2}=1$ and $\lambda_{*}=10$\ Mpc (with $\beta_{*1}=0$). The inclusion of gravitational redshift allows us to exclude GBD and thus break the degeneracy between modified gravity and a fifth force.}  
    \label{fig:CQ_beta1p0_lambda10}
\end{figure}

In Fig.~\ref{fig:CQ_beta1p0_lambda10}, we show the marginalized constraints on $\beta_{*1}$ and $\beta_{*2}$ for a CQ fiducial model with $\beta_{*2}=1$, $\lambda_{*}=10$\ Mpc, a combination that is not excluded by current data, see Ref.~\cite{Wang:2023tjj}. The blue contours correspond to the analysis without the dipole, i.e.~without gravitational redshift, while the red contours include it. As expected, when the dipole is not present, we obtain a perfect elliptic degeneracy between $\beta_{*1}$ and $\beta_{*2}$, which is consistent with the analytical expectation in Eq.~\eqref{eq:Geff_tot} for our chosen time evolution in Eq.~\eqref{eq:beta_time_evol}. This means that even though no gravity modifications were included in the mock data vector, the data (RSD, weak lensing, and CMB) are equally well described using either the (correct) CQ model or the GBD one. 

In the GBD case, the best fit is around $\beta_{*1}=0.84$ and not $\beta_{*1}=1$. This reflects the fact that the impact of a dark fifth force with coupling strength $\beta_{*2}=1$ on the growth of cosmic structure can be mimicked by a modification of gravity with smaller coupling $\beta_1=(\rho_c/\rho)(\delta_c/\delta)^{1/2}\beta_2\simeq (\rho_c/\rho)^{3/2}\beta_2\simeq 0.84\, \beta_2$, as can be seen from Eqs.~\eqref{eq:Geff_GBD}-\eqref{eq:Geff_CQ} with our choice of fiducial cosmological parameters. This degeneracy has important consequences when jointly analyzing RSD, weak lensing, and CMB data: if CDM is subject to additional interactions, such modifications in the dark sector could be incorrectly interpreted as evidence for modified gravity, even though GR remains valid. 

The inclusion of gravitational redshift into the analysis decisively breaks the degeneracy between the two models. Indeed, we can clearly see that the blue contours exclude the case $\beta_{*2}=0$, indicating that a pure modification of gravity no longer fits the data. Even with the dipole, a large portion of the parameter space is still allowed, implying that models with both a dark fifth force and a modification of gravity are not excluded. However, such scenarios involving both kinds of modifications would be disfavoured according to the Occam's razor. An alternative representation of the constraints in polar coordinates is presented in Appendix \ref{app:polar_coords}.

\subsection{Varying the coupling strength} \label{sec:results_varying_beta}
\begin{figure}[t]
    \centering
    \includegraphics[width=0.5\textwidth]{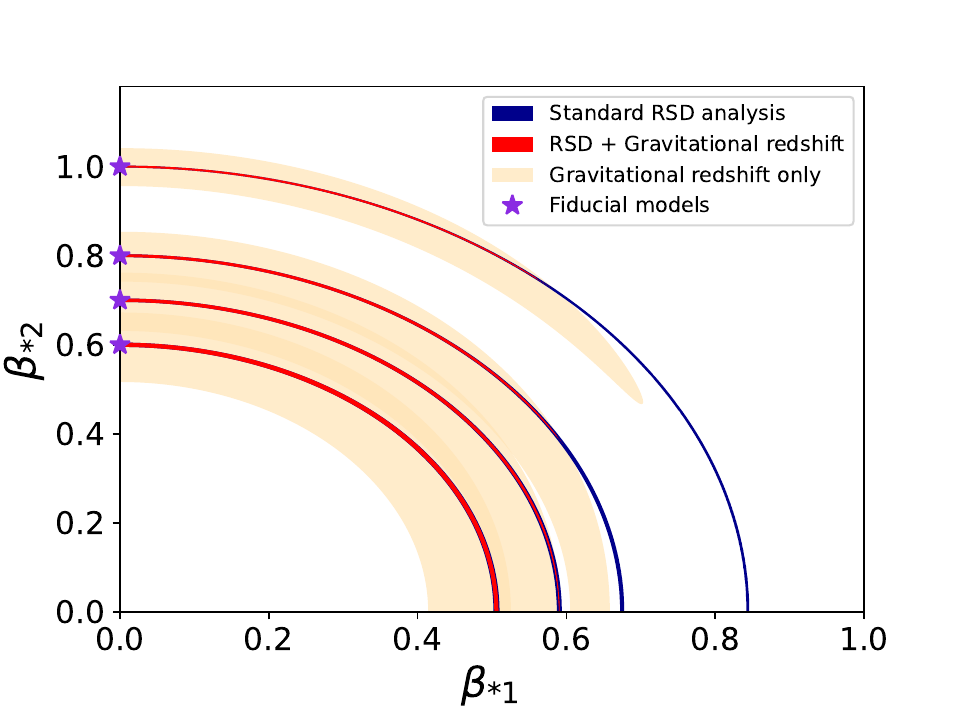}
    \caption{The $1\sigma$ confidence regions for four different fiducial values of the CQ coupling $\beta_{*2}$. For all fiducial models (indicated by stars) we set $\lambda_* = 10$ Mpc. All parameters except for $\beta_{*1}$ and $\beta_{*2}$ have been fixed to their fiducial values.}
    \label{fig:CQ_all_vs_even_all_fids}
\end{figure}

As a next step, we investigate how small the CQ parameter $\beta_{*2}$ can be for the two models to be distinguishable with a SKA2-like survey. In Fig.~\ref{fig:CQ_all_vs_even_all_fids}, we present the constraints on various fiducial models with fixed $\lambda_*=10$\,Mpc and different values of $\beta_{*2}$. To reduce the computational cost of the analysis, we fix the cosmological parameters to their fiducial values, and perform a likelihood minimization analysis only considering the spectroscopic galaxy sample. The CMB and weak lensing data are essential in constraining the cosmological parameters, but the constraints on $\beta_{*1}$ and $\beta_{*2}$ are driven by RSD and gravitational redshift. 

We show the results including the RSD data only (in blue), the dipole only (in yellow) and the combination of the two (in red). Since the cosmological parameters are fixed, the contours are now artificially much tighter than in Fig.~\ref{fig:CQ_beta1p0_lambda10}, but the degeneracy is perfectly captured and we obtain a very good qualitative agreement with Fig.~\ref{fig:CQ_beta1p0_lambda10} for the corresponding fiducial model. We notice that the case $\beta_{*2}=0.7$ is at the edge of the region where the addition of the dipole can exclude a pure GBD scenario, whereas this observable does not give any additional information for the fiducial model with $\beta_{*2}=0.6$. This means that if there is a fifth force acting on CDM with a coupling $\beta_{*2}\leq 0.6$, we will clearly detect the modifications with a survey like SKA2, but we will not be able to determine whether they are due to a dark fifth force or a modification of gravity. On the other hand, if the coupling is larger, the dipole will be able to discriminate between the two scenarios. 

It should be noted that the breaking of the degeneracy between $\beta_{*1}$ and $\beta_{*2}$ does not arise from the total dipole in Eq.~\eqref{eq:dipole}, but only from the part induced by the gravitational redshift, namely, the $P_{\delta \Psi}$ term. It is only this specific term that carries the additional information to distinguish between the two models and determines the lower bound in $\beta_{*2}$ for the dipole to be effective.  

\subsection{Varying the Compton wavelength}

\begin{figure}[t]
    \centering
    \includegraphics[width=0.5\textwidth]{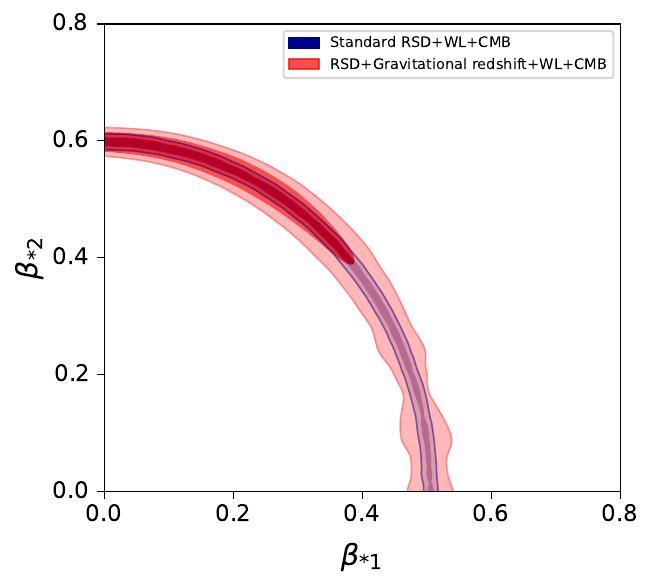}
    \caption{As in Fig.~\ref{fig:CQ_beta1p0_lambda10}, but for fiducial parameters $\beta_{*2}=0.6$ and $\lambda_{*}=100$\ Mpc (with $\beta_{*1}=0$). The inclusion of gravitational redshift allows us to exclude GBD at 1$\sigma$.}
    \label{fig:CQ_beta0p6_lambda100}
\end{figure}

In Figs.~\ref{fig:CQ_beta1p0_lambda10} and \ref{fig:CQ_all_vs_even_all_fids}, we have fixed the value of $\lambda_{*}$ to $10$\,Mpc. Since $\lambda_{*}$ governs the Compton wavelength of the force mediated by the scalar field, a larger value of this parameter would generate modifications in the clustering observables at larger scales. In Fig.~\ref{fig:CQ_beta0p6_lambda100}, we show the constraints on mock data generated for a CQ fiducial model with $\beta_{*2}=0.6$ and $\lambda_{*}=100$\,Mpc, where we see that the dipole is able to break the degeneracy between the two models at the 1$\sigma$ level. This is due to the fact that, for $\lambda_{*}=100$\,Mpc, this observable is significantly modified at large scales when $\beta_{*2}=0.6$, contrary to the previous case with $\lambda_{*}=10$\,Mpc, where the modifications are very small for this value of $\beta_{*2}$.

\subsection{The range of validity of our results}

All the results presented so far concern GR models with a dark fifth force acting on CDM. Since the situation is symmetric, the same conclusions also apply in the case of GBD modified gravity with no CDM interactions. In this case, the elliptic degeneracy without gravitational redshift will be such that mock data generated by a GBD model with $\beta_{*1}=1$ can be well described by a CQ scenario with $\beta_{*2}\simeq (\rho_c/\rho)^{-3/2}\simeq 1.19$. Including gravitational redshift allows us to distinguish between the two scenarios down to $\beta_{*1}\simeq 0.7$ for $\lambda_* = 10$ Mpc. 

Finally, for the dipole to be helpful in this analysis, let us briefly discuss two quantities that impact the aforementioned thresholds in $\beta_{*1}$ and $\beta_{*2}$. First, the bias difference between the correlated galaxy populations directly governs the amplitude of the gravitational redshift contribution, which in turn determines the ability of the dipole to discriminate between the models. Here, we have assumed a bias difference of 1, consistent with the measurements from BOSS~\cite{Gaztanaga:2015jrs}. The population of galaxies detected by SKA2 may significantly differ from that of BOSS, possibly leading to a smaller bias difference. However, it is possible to boost this difference by exploring different ways to divide the galaxies into two populations, for example performing density splits based on their environment, which can increase the amplitude of the gravitational redshift contribution \cite{Paillas:2021oli, Beutler:2020evf}. 

Secondly, the range of scales considered in the analysis also has an influence on the role played by the dipole. Here, we have adopted a minimum separation of $20\,h^{-1}$Mpc, since at such scales nonlinear effects are expected to have a small impact on the dipole in GR~\cite{Bonvin:2023jjq}. The situation may be slightly different in modified gravity. We checked that if the minimum separation is raised to $32\,h^{-1}$Mpc, the constraining power of the dipole slightly decreases, leading to wider contours. As a consequence, the limiting values for $\beta_{*1}$ and $\beta_{*2}$ obtained in Sec.~\ref{sec:results_varying_beta} would become somewhat larger, but the main message of our work would remain unchanged.

\section{Conclusions}\label{sec:conclusions}
Combining RSD with gravitational lensing is generally regarded as the optimal way to test for deviations from GR. Since RSD probes the evolution of structure, while gravitational lensing is sensitive to the geometry of the Universe, their combination would test for the gravitational slip \cite{Amendola:2007rr, Daniel_2009, Zhao:2010dz, Song_2011, Motta:2013cwa}, which in a modified gravity scenario is due to an additional degree of freedom mediating a force on all non-relativistic matter. However, as we have shown in this paper, these two observables cannot distinguish a modification of gravity from additional forces acting on CDM. By performing an MCMC analysis, we have demonstrated that mock data generated by a coupled quintessence model with a fifth force acting on CDM is equally well fitted by a GBD modified gravity scenario. This means that, in the future, we may wrongly claim to have discovered a deviation from GR on cosmological scales, whereas in reality the modifications are due to new interactions in the dark sector (or vice versa). 

This ambiguity can be resolved by including the effect of gravitational redshift into the analysis. Thanks to its sensitivity to the temporal distortion $\Psi$ of the metric, gravitational redshift combined with RSD directly probes the validity of the Euler equation for CDM. This provides an efficient way to discriminate between dark interactions and modifications of gravity. Here, we have quantified the constraining power of gravitational redshift by comparing a GBD modification of gravity with a CQ model, finding that modifications that are still in agreement with current constraints will be distinguishable by a future survey like SKA2. This is particularly remarkable, since measuring gravitational redshift does not require new data, but can be simply achieved by augmenting the standard RSD analysis with the dipole generated in the cross-correlation of different galaxy samples. 

For the purpose of this demonstration, we have considered a scenario where the time evolution of both modifications is exactly the same. This was intentionally chosen so that the two models could be treated on an equal footing and would yield the very same impact on the growth of cosmic structure. A different time evolution could in principle partly break the degeneracy obtained with RSD data, but, since this would not be known {\it a priori} when analyzing real data, we consider scenarios where the two models can exactly mimic each other.

Depending on the size of the deviation from $\Lambda$CDM, the dipole may not be able to distinguish between the GBD and CQ scenarios. However, since this signal is directly related to the bias difference between the two populations of galaxies being correlated, it is worth exploring strategies on how to boost this difference so as to achieve the best possible constraints. This indicates a promising avenue to shed light on the cosmological dark sector in an unambiguous way.

\acknowledgments
SC, LD and CB acknowledge funding from the European Research Council (ERC) under the European Union’s Horizon 2020 research and innovation program (Grant agreement No.~863929; project title ``Testing the law of gravity with novel large-scale structure observables"). CB is also supported by the Swiss National Science Foundation. ZW and LP acknowledge support by the National Sciences and Engineering Research Council of Canada. This research was enabled in part by support provided by the BC DRI Group and the Digital Research Alliance of Canada (\url{alliancecan.ca}). Part of the numerical analysis was completed on the Baobab high-performance computing cluster at the University of Geneva. We gratefully acknowledge the use of {\sc GetDist}~\cite{Lewis:2019xzd}.

\newcounter{appendixsection}
\renewcommand{\theappendixsection}{\Alph{appendixsection}}
\appendix
{\normalfont\normalsize\bfseries}

\stepcounter{appendixsection}
\section{Generalized Brans-Dicke and coupled quintessence scenarios}\label{app:GBD_and_CQ}
Below we provide the relevant equations in the GBD and CQ models, following the notations and conventions of Ref.~\cite{Bonvin:2022tii}. In particular, we work in the baryon frame, i.e.~the frame where Standard Model (SM) particles follow the geodesics of the metric. The action for the GBD model is given by
\begin{equation}\label{eq:GBD_action}
\begin{split}
    S^{\mathrm{GBD}} & = \int {\mathrm{d}}^4 \sqrt{-g} \left[ \frac{A^{-2} (\phi)}{16 \pi G} R - \frac{1}{2} \partial_\mu \phi \, \partial^\mu \phi - V (\phi) \right. \nonumber \\
    &  + \mathcal{L}_{\mathrm{m}} (\psi_{\mathrm{CDM}}, \psi_{\mathrm{SM}}, g_{\mu \nu})  \bigg],
\end{split}
\end{equation}
where $G$ is the Newton constant, $g$ is the determinant of the baryon frame metric $g_{\mu \nu}$ and $R$ is the associated Ricci scalar. The scalar field is denoted with $\phi$, $V$ is its potential and $A$ is a generic function of $\phi$. The contributions of SM and CDM particles, denoted by $\psi_{\mathrm{SM}}$ and $\psi_{\mathrm{CDM}}$ respectively, are included in the matter Lagrangian density $\mathcal{L}_{\mathrm{m}}$. The scalar field is conformally coupled to all matter, such that also CDM follows the geodesics of the baryon frame metric $g_{\mu \nu}$.

On the other hand, in the CQ scenario, the scalar field is conformally coupled to CDM only. In the baryon frame, the action is given by
\begin{equation}
\begin{split}
    S^{\mathrm{CQ}} & = \int {\mathrm{d}}^4 \sqrt{-g} \left[ \frac{1}{16 \pi G} R - \frac{1}{2} \partial_\mu \phi \, \partial^\mu \phi - V (\phi) \right. \nonumber \\
    &  + \mathcal{L}_{\mathrm{SM}} (\psi_{\mathrm{SM}}, g_{\mu \nu}) + \mathcal{L}_{\mathrm{CDM}} (\psi_{\mathrm{CDM}}, A^2(\phi)g_{\mu \nu}) \bigg]\, ,
\end{split}
\end{equation}
where the gravitational part of the action is not modified and CDM particles follow geodesics of $A^2(\phi)g_{\mu \nu}$.

We adopt the line element in Eq.~\eqref{eq:perturbed_FLRW} and work under the quasi-static approximation, neglecting time derivatives of the metric and the field perturbations over spatial ones~\cite{Mirpoorian:2023utj}. For GBD, this leads to the following sets of equations in Fourier space on subhorizon scales: 
\begin{align}
&k^2 \Phi = -4 \pi Ga^2 \left(\rho_{b} \delta_{b} + \rho_{c} \delta_{c}\right) - \beta_1 k^2 \delta \phi \label{eq:poisson_st} \\
&k^2 (\Phi - \Psi) = -2 \beta_1 k^2 \delta \phi \label{eq:slip} \\
&\dot{\delta}_{b}+\theta_{b} = 0 \label{eq:STcont_b} \\
&\dot{\theta}_{b}+{\cal H} \theta_{b} = k^{2} \Psi \label{eq:STeuler_b}\\
&\dot{\delta}_{c}+\theta_{c} = 0 \label{eq:STcont_c}\\
&\dot{\theta}_{c}+{\cal H} \theta_{c} = k^{2} \Psi \label{eq:STeuler_c}\\
\label{eq:dphi}
&\delta \phi = - \frac{\beta_1 (\rho_{c} \delta_{c} + \rho_{b} \delta_{b})}{m^2 + k^2/a^2} \\
&\square \phi = V_{, \phi}+\beta_1({\rho_{c}+\rho_b}) \equiv V^{\rm eff},_{\phi} \label{eq:Veff_st}\\
\label{eq:delta_ddot_st}
&\ddot{\delta}+ {\cal H} \dot{\delta} = 4 \pi G a^2 \rho\delta\!\left[ 1 + {2 \tilde{\beta_1}^2k^2\over a^2m_1^2 + k^2} \right].
\end{align}
Here, $\Box \equiv \nabla^{\mu} \nabla_{\mu}$, the scalar field coupling strength is given by $\beta_1 = A_{, \phi}/{A}$, where a comma indicates a derivative, and we have defined $\tilde{\beta}_1\equiv\beta_1/\sqrt{8\pi G}$.

For CQ, the analogous set of equations is
\begin{align}
&k^2 \Phi = -4 \pi Ga^2 \left(\rho_{b} \delta_{b} + \rho_{c} \delta_{c} \right) \\
&k^2 (\Phi - \Psi) = 0 \\
&\dot{\delta}_{b}+\theta_{b} = 0 \label{eq:CQcont_b} \\
&\dot{\theta}_{b}+ {\cal H} \theta_{b} = k^{2} \Psi \label{eq:CQeuler_b} \\
&\dot{\delta}_{c}+\theta_{c} = 0 \label{eq:CQcont_c}\\
&\dot{\theta}_{c}+ ({\cal H} +\beta_2 \dot{\phi} )\theta_{c} = k^{2} \Psi + k^{2} \beta_2 \delta \phi \label{eq:CQeuler_c}\\
\label{eq:dphi_cq}
&\delta \phi = - \frac{\beta_2 \rho_c \delta_c}{m_2^2 + k^2/a^2} \\
\label{eq:Veff_cq}
&\square \phi = V_{, \phi}+\beta_2 \rho_{c} \equiv V^{\rm eff},_{\phi} \\
&\ddot{\delta}+ \HH \dot{\delta} = 4 \pi G a^2 \rho\delta\!\left[ 1 + {2 \tilde{\beta_2}^2k^2\over a^2m^2 + k^2} \left( \rho_c \over \rho \right)^2\!\! \left( \delta_c \over \delta \right) \right],
\label{eq:delta_ddot_cq}
\end{align}
where $\tilde{\beta}_2$ is defined in the same way as $\tilde{\beta}_1$. The effective potential $V^{\rm eff}$ is defined via Eqs.~\eqref{eq:Veff_st} and \eqref{eq:Veff_cq} for the two scenarios and related to the scalar field mass via $m^2 = V^{\rm eff}_{, \phi \phi}$. 

Following Ref.~\cite{Bonvin:2023jjq}, we assume $A^{-2} \approx 1$, with no impact on the arguments presented in our analysis. For GBD, this implies that the Newton constant $G$ appearing in the equations always corresponds to the present-time value, which is a robust assumption since the redshift evolution of the gravitational coupling is constrained to be very small in screened GBD theories \cite{Wang:2012kj}. In the case of CQ, allowing for $A^{-2} \neq 1$ simply corresponds to a rescaling of the coupling $\beta_2$.

\section{Galaxy clustering survey specifications}
\label{app:specifications}

We adopt the SKA2 specifications from Ref.~\cite{Bull:2015lja}, assuming measurements of the clustering correlation function multipoles in 15 spectroscopic redshift bins centered at $z=0.15,0.25,\ldots,1.55$, each with width $\Delta z=0.1$. In each bin, we evaluate the multipoles at 35 separations from $d=20\,h^{-1}$Mpc to $d=160\,h^{-1}$Mpc in increments of $4\,h^{-1}$Mpc. 

We model the galaxy bias according to the fitting function from Ref.~\cite{Bull:2015lja},
\begin{equation}
b_\mathrm{P}(z) = b_\mathrm{P,1}\exp(b_\mathrm{P,2}\:\! z)\pm\Delta b/2,
\end{equation}
where P indicates the bright (B) or faint (F) galaxy population. We let the parameters $b_\mathrm{P,1}$ and $b_\mathrm{P,2}$ free with fiducial values $0.554$ and $0.873$, respectively, and we set $\Delta b=1$ in agreement with the bias difference measured in Ref.~\cite{Gaztanaga:2015jrs}. 

For the purpose of our forecasts, we fix the magnification bias according to Appendix B in Ref.~\cite{Castello:2022uuu} and set the evolution bias to 0. Both quantities will be directly measurable from the average number of galaxies once the data become available. \\

\section{Constraints in polar coordinates}\label{app:polar_coords}

An alternative visualization of the constraints can be achieved by performing a change to polar coordinates, 
\begin{align}
    R  = \sqrt{\beta_{*1}^2 + \beta_{*2}^2}\,,\qquad
    \theta  = \frac{\beta_{*2}}{R}\,.
\end{align} 

\begin{figure}[h]
    \centering
    \includegraphics[width=0.5\textwidth]{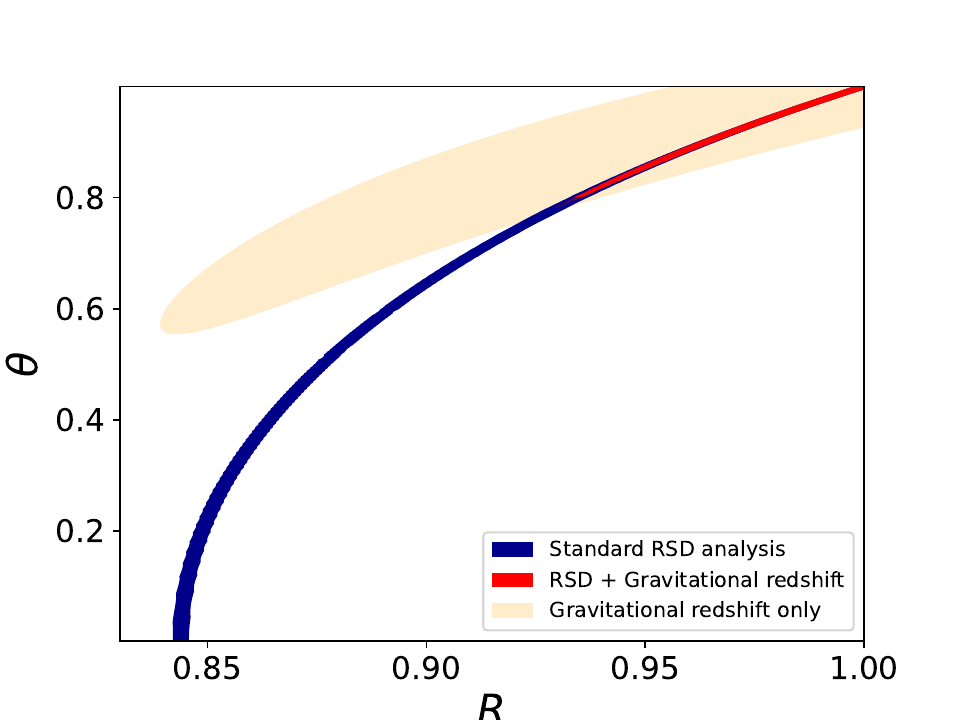}
    \caption{The $1\sigma$ and confidence regions on the polar coordinates $R$ and $\theta$. Here, the fiducial model is CQ and is specified by $\beta_{*2}=1$ and $\lambda_{*}=10$\ Mpc (with $\beta_{*1}=0$).}  
    \label{fig:polar_coords}
\end{figure}

We show in Fig.~\ref{fig:polar_coords} the 1$\sigma$ confidence regions on $R$, $\theta$ for the same CQ fiducial model as in Fig.~\ref{fig:CQ_beta1p0_lambda10}, i.e.~with $\beta_{*2} = 1$, $\lambda_* = 10$ Mpc. For simplicity, these constraints were obtained by fixing all cosmological parameters to their fiducial values.

\stepcounter{appendixsection}

\bibliography{Einstein_Euler_forecasts}
\bibliographystyle{apsrev4-1}

\end{document}